
\magnification=1200
\hsize=13cm
\centerline{\bf 1. Introduction}
\vskip 1cm
Anyons are two-dimensional particles with arbitrary statistics
interpolating between bosons and fermions [1,2] (for reviews
see for example [3,4] and references therein).
In the past few years they have attracted considerable
interest especially in connection with the interpretation
of certain condensed matter phenomena, most notably
the fractional quantum Hall effect [5]. Since the first
seminal papers on the subject [1,6], it was clear that anyons
are deeply connected with the braid group of which they are
abelian representations, just like bosons and fermions are
abelian representations of the permutation group. The appearance
of the braid group in place of the permutation group is a peculiar
feature of two dimensional systems; in fact as is well known,
in three or more dimensions only bosons and fermions can exist
so that the manybody wavefunctions are either symmetric or antisymmetric
in the exchange of any two identical particles. In two dimensions
instead,
when one exchanges two identical particles, it is no longer enough to
compare their final ordering to the original one ({\it i.e.} to specify
their permutation) but it is necessary to specify also how the
exchanging trajectories of the two particles wind or braid around each
other. These braiding properties make the quantum mechanics of anyons
extremely difficult.

One possibility to shed some light on this problem could be
to explore and study the characteristic symmetries of anyon
systems. The natural candidates for these appear to be the
quantum groups \footnote{${}^1$}{Here and in the following,
we adopt the commonly used terminology. However more properly,
one should speak of quantum enveloping algebras.}, which are
deformations of ordinary Lie algebras [7-9]. One of the fundamental
features of quantum groups is that their centralizer is the braid
group, just like the permutation group is the centralizer of the
ordinary Lie algebras . In other words quantum groups are endowed
with a comultiplication \footnote{${}^2$}{The comultiplication is
the operation which is used to make tensor products of representations.}
which is not cocommutative but envolves suitable braiding factors [7-13].

The fundamental role played by the braid group both in the theory
of quantum groups and in the theory of anyons suggests that a deep
relation between these two subjects might exist.
In this paper we show that this is indeed the case.
It is well known that bosonic or fermionic oscillators,
characterized by commutative or anti-commutative Heisenberg algebras,
can be combined \`a la Schwinger to construct non-abelian Lie algebras
 with the permutation group as their centralizer [14]. Similarly one can
think of using anyonic oscillators with braid group properties and
$q$-deformed commutation relations to build non-abelian algebras with
the braid group as their centralizer, {\it i.e.} to construct non-abelian
quantum groups. So far, this connection between anyons and quantum
groups
has not been investigated in the literature despite intensive studies
in both fields. In this paper we elaborate on this idea and show
how to construct quantum groups from anyons.

Actually, oscillators with $q$-deformed commutation relations have
been introduced a few years ago and are known as $q$-oscillators
[15-18]. Later, the standard bosonic and fermionic Schwinger
constructions
for $SU(2)$ have been generalized to $q$-oscillators yielding the
quantum
group $SU(2)_q$. We would like to stress that despite the many formal
analogies, our anyonic construction does not have anything to do with
that of the $q$-oscillators. This is so for several reasons. First of
all,
the $q$-oscillators can be defined in any dimensions and are not related
to the braid group, whereas anyons are strictly two dimensional objects.
Secondly, the $q$-oscillators are local operators, whilst anyons are
intrinsically {\it non-local} due to their braiding properties [19].
This non-locality, which is essential to distinguish whether anyons
are exchanged clockwise or anticlockwise, allows also to define a
natural ordering among the particles, which in turn is essential to
define a non-cocommutative comultiplication like the one of quantum
groups.

This paper is organized as follows. In section 2, we rewiew the
standard bosonic and fermionic Schwinger construction of $SU(2)$ and
briefly discuss its $q$-oscillator extension. In section 3, we
introduce anyonic oscillators of statistics $\nu$ on a two dimensional
square lattice by means of the Jordan-Wigner construction [20],
and extensively discuss their generalized commutation relations.
The choice of working on a lattice is dictated by the need of having a
discrete set of particles so that later we will be able to define
a discrete comultiplication with no ambiguitites. However,
to have simple
deformed commutation relations for the anyonic oscillators, it is
convenient to take a sort of a continuum limit that can be realized
by embedding the original lattice into one with an infinitesimal
spacing.
In section 4, we use the anyonic oscillators of statistics $\nu$ to
construct the generators of the quantum group $SU(2)_q$ where the
deformation parameter is $q=\exp({\rm i}\pi\nu)$. Finally, in section 5
we present our conclusions.

\vskip 4cm
\centerline{\bf 2. Bosonic and Fermionic Constructions of $SU(2)$}
\vskip 1cm
It is very well known that the $SU(2)$ algebra can be explicitly
realized in several ways. One of the simplest of these is the
Schwinger construction [14]. Given a pair of bosonic
harmonic oscillators such that
$$
\left[a_i~,~a_j^\dagger\right] ~=~ \delta_{ij}
\eqno(2.1)
$$
for $i,j=1,2$, the three generators of $SU(2)$ are realized as
bilinears in $a$ and $a^\dagger$ according to
$$
\eqalign{
j^+ &= a_1^\dagger ~a_2 \ \ ,\cr
j^0 &= {1\over 2}\left(a_1^\dagger~a_1 - a_2^\dagger~a_2\right)
\ \ ,\cr
j^- &= a_2^\dagger ~a_1 \ \ .}
\eqno(2.2)
$$
Indeed, using (2.1) it follows immediately that
$$
\eqalign{
\left[j^+~,~j^-\right]~&=~2j^0 \ \ ,\cr
\left[j^0~,~j^\pm\right]~&=~\pm j^\pm \ \ .}
\eqno(2.3)
$$
It is interesting to point out that using a {\it single} pair of
bosonic oscillators we can obtain not only an algebraic realization
of $SU(2)$ but also the full set of its (unitary) representations.
To see this, let us define the (normalized) states
$$
|j,m\rangle = {1\over {\sqrt{(j+m)!\,(j-m)!}}}
\left(a_1^\dagger\right)^{j+m} \left(
a_2^\dagger\right)^{j-m} |0\rangle
\eqno(2.4)
$$
where $j$ and $m$ are arbitrary integers or half-integers such that
$j\pm m \in {\bf Z}_+\cup \{0\}$, and the state $|0\rangle$ is
the ``vacuum'' which satisfies
$$
a_i\,|0\rangle = 0
\eqno(2.5)
$$
for $i=1,2$. Using (2.1), it is easy to verify that
$$
\eqalign{
j^0~|j,m\rangle ~&=~ m\,|j,m\rangle \ \ , \cr
j^\pm~|j,m\rangle ~&=~ \sqrt{(j\mp m)(j\pm m +1)}\,
|j,m\pm1\rangle \ \ .}
\eqno(2.6)
$$
Thus $|j,m\rangle$ are the familiar angular momentum states in
which $j$ labels the
``total'' spin and $m$ its third component. Since $j$
can take arbitrary positive integer or half-integer values,
all possible unitary representations of $SU(2)$ are spanned by
the states (2.4) on which the corresponding generators are
simply given by (2.2).

The Schwinger construction can be realized also using fermions
instead of bosons. In fact let us consider a pair of fermionic
oscillators obeying the following anticommutation relations
$$
\left\{c_i~,~c_j^\dagger\right\} =
\delta_{ij}
\eqno(2.7)
$$
for $i,j=1,2$. Then, in analogy with (2.2) let us define
$$
\eqalign{
j^+ &= c_1^\dagger ~c_2 \ \ ,\cr
j^0 &= {1\over 2}\left(c_1^\dagger~c_1 -
c_2^\dagger~c_2\right) \ \ ,\cr
j^- &= c_2^\dagger ~c_1 \ \ .}
\eqno(2.8)
$$
It is again straightforward to verify that these generators
indeed close the $SU(2)$ algebra.
However, in contrast with the bosonic case it is now impossible
to recover the full set of the unitary representations of $SU(2)$.
In fact, denoting by $|0\rangle$ the fermionic vacuum
($c_i\,|0\rangle=0$
for $i=1,2$), we can construct only the following four states
$$
|0\rangle~~~,~~~c_1^\dagger|0\rangle~~~,~~~
c_2^\dagger|0\rangle~~~,~~~
c_1^\dagger c_2^\dagger|0\rangle~~~,
\eqno(2.9)
$$
because the anticommutation relations prevent to put more
than one fermion of each kind, and
thus only the $j=0$ and $j=1/2$ representations are realized.
To overcome this problem we must consider many copies of pairs
of fermionic oscillators, or more properly we must
consider a two-component Pauli spinor field
$$
c(x) =
\left(\matrix{
 c_1(x) \cr
 c_2(x) \cr}\right)
\eqno(2.10)
$$
where $x$ belongs to some manifold $\Omega$ to be specified
later (we can think of
$x$ as the label that distinguishes the different copies of
the fermionic oscillators).
The adjoint of $c(x)$ is given by
$$
c^\dagger(x) =\left( c_1^\dagger(x)~,~c_2^\dagger(x)\right)
\eqno(2.11)
$$
where $c^\dagger_1(x)$ can be interpreted as the operator
creating
a fermion of type 1 (``spin up'') at the point $x \in \Omega$, and
$c^\dagger_2(x)$ as the operator creating a fermion of type 2
(``spin down'') at $x$. Moreover we assume the standard
anticommutation relations, {\it i.e.}
$$
\left\{c_i(x)~,~c_j^\dagger(y)\right\} = \delta_{ij}~\delta(x,y)
\eqno(2.12)
$$
where $\delta(x,y)$ is the delta function on $\Omega$.
Then we can define the local operators
$$
\eqalign{
j^+(x) &= c_1^\dagger(x) c_2(x) \ \ , \cr
j^0(x) &= {1\over 2}\left(c_1^\dagger(x)c_1(x) -
c_2^\dagger(x)c_2(x)\right) \ \ , \cr
j^-(x) &= c_2^\dagger(x) c_1(x) \ \ , }
\eqno(2.13)
$$
which obey the following commutation relations
$$
\eqalign{
\left[j^+(x)~,~j^-(y)\right]~&=~2j^0(x)~\delta(x,y) \ \ ,\cr
\left[j^0(x)~,~j^\pm(y)\right]~&=~\pm j^\pm(x)~\delta(x,y) \ \ .}
\eqno(2.14)
$$
Eqs. (2.14) indicate that a ``local'' $SU(2)$ algebra is
realized at each point of $\Omega$. Obviously these local
algebras have only
the spin 0 and spin 1/2 representations which we may call
``local'' representations.

A global algebra can be readily constructed by combining local
ones with a repeated use of comultiplication. In fact using (2.14),
it is easy to check that the operators
$$
\eqalign{
J^\pm&\equiv\sum_x J^\pm(x) =\sum_x {\bf 1}\otimes\cdots\otimes
{\bf 1}\otimes
j^\pm(x)\otimes{\bf 1}\otimes\cdots{\bf 1} \ \ ,\cr
J^0&\equiv\sum_x J^0(x) =\sum_x {\bf 1}\otimes\cdots
\otimes{\bf 1}\otimes
j^0(x)\otimes{\bf 1}\otimes\cdots{\bf 1} }
\eqno(2.15)
$$
generate a global $SU(2)$ algebra. The symbol $\otimes$ in (2.15)
denotes the direct product
so that the operators $J^\pm(x)$ and $J^0(x)$ act as the identity
at all points other
than $x$ and as $j^\pm(x)$ and $j^0(x)$, respectively, at $x$.
By combining the local spin 0 and spin 1/2 representations it
is possible to construct
{\it all} the unitary representations of the global $SU(2)$ algebra.
For example the spin
1 representation is carried by the space spanned by the following
three states
$$
c_1^\dagger(x_1)c_1^\dagger(x_2)|0\rangle~~~,~~~
{1\over{\sqrt 2}}\left(c_1^\dagger(x_1)c_2^\dagger(x_2)+
c_2^\dagger(x_1)c_1^\dagger(x_2)\right)|0\rangle~~~,~~~
c_2^\dagger(x_1)c_2^\dagger(x_2)|0\rangle\ \ ,
$$
where $x_1$ and $x_2$ are two arbitrary distinct points in $\Omega$.
All other representations
can be realized in a similar way.

The Schwinger construction of $SU(2)$ has been recently
generalized to the so-called $q$-oscillators [15-18]. These are
deformations of the ordinary harmonic oscillators
characterized by the following generalized commutation relations
$$
{\tilde a}_i\,{\tilde a}^\dagger_i - q^{-1}\,
{\tilde a}^\dagger_i\,{\tilde a}_i=q^{{\tilde N}_i} \ \ ,
\eqno(2.16)
$$
where $q$ is the deformation parameter and ${\tilde N}_i$ is
the number operator. Notice that ${\tilde N}_i$
is not ${\tilde a}^\dagger_i\,{\tilde a}_i$, but it
nevertheless satisfies the usual relations with ${\tilde a}_i$
and ${\tilde a}^\dagger_i$, namely
$$
\Big[{\tilde N}_i~,~{\tilde a}^\dagger_i\Big]=
{\tilde a}^\dagger_i~~~~~,~~~~~
\Big[{\tilde N}_i~,~{\tilde a}_i\Big]=
-{\tilde a}_i\ \ .
\eqno(2.17)
$$
Only in the limit $q\to 1$, we retrieve the standard
bosonic Heisenberg algebra of the harmonic oscillator
and ${\tilde N}_i$ becomes ${\tilde a}^\dagger_i
\,{\tilde a}_i$.
Using ${\tilde a}_i$ and ${\tilde a}^\dagger_i$ for
$i=1,2$ in the Schwinger approach, one can define the
following operators
$$
\eqalign{
J^+&={\tilde a}^\dagger_1\,{\tilde a}_2\ \ , \cr
J^-&={\tilde a}^\dagger_2\,{\tilde a}_1\ \ , \cr
J^0&={1\over 2}\left({\tilde N}_1-{\tilde N}_2\right)\ \ , }
\eqno(2.18)
$$
and check that they satisfy [15,16]
$$
\eqalign{
\left[J^0~,~J^\pm\right]&=\pm J^\pm \ \ ,\cr
\left[J^+~,~J^-\right]&={{q^{2J^0}-q^{-2J^0}}
\over{q-q^{-1}}} \ \ .}
\eqno(2.19)
$$
These are the commutators of the quantum group $SU(2)_q$
[7-9], and thus one can say that the Schwinger construction for
$q$-oscillators naturally leads to a quantum group with $q$
as deformation parameter.

It is worthwhile to mention that the standard Schwinger
construction of $SU(2)$ can be generalized also in a different
way by using fermionic oscillators together with a non
cocommutative comultiplication [11]. For this to be possible
however, it is necessary to order the fermions for example by
putting them on a line. Then one can define
$$
\eqalign{
J^\pm_q&=\sum_{x}J^\pm_q(x)=\sum_x
\left(\prod_{ y< x}q^{-2j^0(y)}~
j^\pm(x)~\prod_{z>x}
q^{2j^0(z)}\right)\ \ , \cr
J^0_q&=\sum_xJ^0_q(x)=\sum_xj^0(x) \ \ ,}
\eqno(2.20)
$$
where $j^\pm(x)$ and $j^0(x)$ are given by eq. (2.13), $x$
is the coordinate of a one dimensional chain and $q$ is an
arbitrary complex number. From the local algebra
(2.14) it is easy to check that
$$
\eqalign{
\left[J^0_q~,~J^\pm_q\right]&=\pm J^\pm_q \ \ ,\cr
\left[J^+_q~,~J^-_q\right]&={{q^{2J^0_q}-q^{-2J^0_q}}
\over{q-q^{-1}}} \ \ .}
\eqno(2.21)
$$
These are again the commutators of $SU(2)_q$. It is interesting to
observe that since $\left(j^+(x)\right)^\dagger=j^-(x)$ and
$\left(j^0(x)\right)^\dagger=j^0(x)$, we have
$$
\left[J^+_q(x)\right]^\dagger=J^-_{q^\star}(x)
\eqno(2.22)
$$
for any $q$. This construction appears naturally in several
one-dimensional quantum
spin systems (like for example the XXZ model) of which $J^\pm_q$
and $J^0_q$ turn out to be symmetry operators [11].

In the next sections we will present a construction of $SU(2)_q$
which, even though still inspired by the Schwinger approach, is
completely different from the ones we have just mentioned.
Indeed our construction will exploit {\it anyonic} operators,
which contrarily to the $q$-oscillators, carry a representation
of the braid group and are intrinsically non-local objects.
Moreover, in our case the non cocommutativity of the comultiplication
will be an automatic consequence of the statistics of
the anyonic operators.

\vskip 4cm
\centerline{\bf 3. Lattice Angle Function and Anyonic Oscillators}
\vskip 1cm
In the following two sections we are going to generalize
the Schwinger construction to the case
of anyonic oscillators of statistics $\nu$ which
continuously
interpolate between bosons and fermions. As is well known,
anyons can exist only in two space dimensions where the braid
group replaces the permutation group in the classification of all
possible statistics [4]. Therefore despite of the many
formal analogies betweem them, anyonic oscillators do not have to
be confused with the $q$-oscillators mentioned in the previous section
which in principle can exist in any dimension. Since we will be
interested in the anyonic case,
from now on we will work only with objects defined on a two
dimensional manifold $\Omega$ whose points will be denoted
by ${\bf x} = (x_1,x_2)$. Moreover for reasons which will be
clear later, we take $\Omega$ to be a two dimensional lattice
(for definiteness a square
lattice) with spacing $a=1$.

The first step of our analysis is the construction of
anyonic oscillators on $\Omega$. Several recent papers have
already analyzed this problem [21-26], but nevertheless we
are going to review it again to set the notations and above
all to point out a few important subtleties that have been
overlooked in the literature.
Our general strategy is to implement on the lattice $\Omega$
the Jordan-Wigner transformation
[20] which allows to transmute for example fermions into
bosons in any dimension and fermions into anyons of arbitrary
statistics in two space dimensions. We remark that in this case
the Jordan-Wigner transformation is inspired by the Chern-Simons
construction of anyons [2,21,27,28] to which it is intrinsecally
related. An essential ingredient of such transformation is the
so called angle function. In the continuous plane ${\bf R}^2$
the angle function is a rather familiar object. Formally it
can be defined  through the
Green function of the Laplace operator, {\it i.e.} through
the function
$$
G({\bf x},{\bf y}) = \ln |{\bf x}-{\bf y}|
\eqno(3.1)
$$
which satisfies
$$
{\partial \over {\partial x^i}}{\partial \over
{\partial x^i}}G({\bf x},{\bf y})=2\pi
\,\delta({\bf x}-{\bf y})\ \ .
\eqno(3.2)
$$
Then, if we introduce the vector field
${\bf f}({\bf x},{\bf y})=\left(
f^1({\bf x},{\bf y}),f^2({\bf x},
{\bf y})\right)$ according to
$$
f^i({\bf x},{\bf y})= - \epsilon^{ij}
{\partial \over {\partial x^j}}G({\bf x},{\bf y})
= - \epsilon^{ij}{x_j\over{|{\bf x}-{\bf y}|^2}}
\eqno(3.3)
$$
where $\epsilon^{ij}$ is the completely antisymmetric
symbol ($\epsilon^{12}=-\epsilon^{21}=1$), the angle
function $\Theta({\bf x},{\bf y})$ is defined by
$$
{\partial \over {\partial x^i}}\Theta({\bf x},{\bf y})=
f_i({\bf x},{\bf y})\ \ ,
\eqno(3.4)
$$
and satisfies
$$
\epsilon^{ij}{\partial \over {\partial x^i}}{\partial
\over {\partial x^j}}
\Theta({\bf x},{\bf y}) = 2\pi \,\delta({\bf x}-
{\bf y})\ \ .
\eqno(3.5)
$$
A solution of eq. (3.4) is
$$
\Theta({\bf x},{\bf y}) = \tan^{-1}\left({{y_2-x_2}
\over{y_1-x_1}}\right)\ \ ,
\eqno(3.6)
$$
which is indeed the naive definition of the angle between
two points measured from the positive $x$-axis. In this
formula $x_i$ and $y_i$ are on the same footing (the right hand
side of eq. (3.6) is indeed symmetric under $x_i \leftrightarrow y_i$)
while they are not in eq. (3.4); therefore to remove the ambiguity
one has to say {\it e.g.} that $\Theta({\bf x},{\bf y})$ is the
angle under which
the point ${\bf x}$ is seen from ${\bf y}$. Furthermore,
the function $\Theta$ is multivalued, and hence a cut
has to be chosen. For example one can take $-\pi\le
\Theta({\bf x},{\bf y})<\pi$. With this choice it is not
difficult to verify that
$$
\Theta({\bf x},{\bf y}) - \Theta({\bf y},{\bf x}) =
\left\{\matrix{
 \pi \,{\rm sgn}(x_2-y_2) ~~~~~{\rm for}~~~
x_2\not=y_2 \ \ , \cr
 \pi \,{\rm sgn}(x_1-y_1) ~~~~~{\rm for}~~~
x_2=y_2 \ \ . \cr}\right.
\eqno(3.7)
$$

The angle function can be defined unambiguously also on
a two-dimensional lattice; however some care must be used
in this generalization [22,23,25,26].
First of all, let us recall that there are two lattice
derivative operators, $\partial_i$
and $\tilde\partial_i$, which for any function $f(\bf x)$
are defined
through
$$
\eqalign{
\partial_i f(\bf x)& = f({\bf x} +\hat {\bf i}) -
f({\bf x}) \ \ ,\cr
\tilde\partial_i f({\bf x})& = f({\bf x}) - f({\bf x} -
\hat {\bf i}) \ \ ,\cr
}\eqno(3.8)
$$
where $\hat{\bf i}$ is the unit vector in the positive
$i$-direction ($i=1,2$). In terms
of $\partial_i$ and $\tilde\partial_i$ the correct
lattice version of eq. (3.2) turns out to be
$$
\partial_i \tilde\partial_i~G({\bf x},{\bf y})=
2\pi \,\delta({\bf x},{\bf y})\ \ ,
\eqno(3.9)
$$
where $\delta({\bf x},{\bf y})$ is the lattice
delta function ($\delta({\bf x},{\bf y})=0 $ if
${\bf x}\not ={\bf y}$; $\delta({\bf x},{\bf y})=1
$ if ${\bf x} ={\bf y}$).
The solution of eq. (3.9) is explicitly given
by
$$
G({\bf x},{\bf y}) = {1\over 2\pi}
\int_{-\pi}^{\pi} d^2p~
{\left[1-\cos\,{\bf p}\cdot ({\bf x}-{\bf y})\right]
\over{
2\sum\limits_{i=1}^2(1-\cos p_i)}}\ \ .
\eqno(3.10)
$$
This is a function only of the difference $({\bf x}-{\bf y})$
and can be regarded as the lattice version of the
continuum
Green function given in eq. (3.1). Thus, to define the
lattice angle function we can repeat steps similar to those
in eqs. (3.3-6). We first define the vector field
${\bf f}({\bf x},{\bf y})$ as
$$
f^i({\bf x},{\bf y})= - \epsilon^{ij}\tilde\partial_j
G({\bf x},{\bf y})\ \ ;
\eqno(3.11)
$$
then from eq. (3.9) it follows that
$$
\eqalign{
f^2({\bf x}+{\hat{\bf 1}},{\bf y})-
f^2({\bf x},{\bf y})-&
f^1({\bf x}+{\hat{\bf 2}},{\bf y})+
f^1({\bf x},{\bf y}) \cr
&=\epsilon^{ij}
\partial_i~f^j({\bf x},{\bf y})=
\partial_i \tilde\partial_i~G({\bf x},{\bf y}) =2\pi\,
\delta({\bf x},{\bf y}) \ \ .}
\eqno(3.12)
$$
One possible representation of ${\bf f}$ satisfying (3.12)
is provided by
$$
f^i({\bf x},{\bf y})= \varphi({\bf x},{\bf y}^\star,{\bf x}+
{\hat {\bf i}})
\eqno(3.13)
$$
where $\varphi({\bf x},{\bf y}^\star,{\bf x}+{\hat {\bf i}})$
is the angle under which the oriented link between ${\bf x}$
and ${\bf x}+{\hat {\bf i}}$ is seen from the point
${\bf y}^\star=(y_1+1/2;y_2+1/2)$ as shown in Fig. 1.
Notice that the point ${\bf y}^\star$ belongs to the
dual lattice $\Omega^\star$. With this choice,
$f^i({\bf x},{\bf y})$ is unambiguously defined also
when ${\bf x}={\bf y}$;
on the other hand if ${\bf x}$ and ${\bf y}$ are very
far apart from each other the difference between ${\bf y}$
and ${\bf y}^\star$ becomes negligible; moreover using the
representation (3.13), we can easily realize that
$$
f^i({\bf x},{\bf y})\longrightarrow 0 ~~~{\rm if}~~~
|{\bf x}-{\bf y}|\to \infty \ \ .
\eqno(3.14)
$$
Let us now return to the general properties of ${\bf f}$.
Recalling that $f^i({\bf x},{\bf y})$ is defined on the lattice
link between ${\bf x}$
and ${\bf x}+{\hat{\bf i}}$, we can rewrite eq. (3.12) as
$$
\oint_{\Gamma_x} {\bf f}({\bf x},{\bf y}) = 2\pi\,
\delta({\bf x},{\bf y})
\eqno(3.15)
$$
where $\Gamma_x$ is the positively oriented boundary
of the elementary plaquette
$A_x$ whose lower left corner is ${\bf x}$ (see Fig. 2).
Thus, for any closed curve $\Gamma$ encircling $k$ times
the point ${\bf y}^\star$, we have
$$
\oint_{\Gamma}{\bf f}({\bf x},{\bf y}) = 2\pi k \ \ .
\eqno(3.16)
$$
Given the vector field ${\bf f}$, it is possible to define
unambiguously the angle between two lattice points ${\bf x}$
and ${\bf y}$. To this aim let us consider a path ${\cal P}_x$
which, following the lattice bonds, starts from a base point $B$
(eventually moved to infinity) and ends at the point ${\bf x}$;
then the function
$$
\Theta_{{\cal P}_x}({\bf x},{\bf y}) = \int_{{\cal P}_x}
{\bf f}({\bf x},{\bf y})
\eqno(3.17)
$$
is the lattice angle function. If we use the explicit
representation of ${\bf f}$ given in eq. (3.13), we can describe
$\Theta_{{\cal P}_x}({\bf x},{\bf y})$ as the angle between
the base point $B$ and the point ${\bf x}$ measured from the
point ${\bf y}^\star\in\Omega^\star$ along the curve ${\cal P}_x$
as shown in Fig. 3. This function has all the properties of
any angle.
In fact we easily see that
$$
\epsilon^{ij} \partial_i \partial_j~
\Theta_{{\cal P}_x}({\bf x},{\bf y})
=2\pi\,\delta({\bf x},{\bf y}) \ \ ,
\eqno(3.18)
$$
which is the lattice analogue of eq. (3.5);
moreover it is multivalued because if the path
${\cal P}_x$ winds counterclockwise around the point
${\bf y}^\star$,
$\Theta_{{\cal P}_x}({\bf x},{\bf y})$ increases of
$2\pi$ like any angle centered in ${\bf y}^\star$. Actually,
using eq. (3.16) we have
$$
\Theta_{{\cal P}_x}({\bf x},{\bf y}) -
\Theta_{{{\cal P}'}_x}({\bf x},{\bf y}) =
\oint_{{{\cal P}_x}{{{\cal P}'}_x^{-1}}}
{\bf f}({\bf x},{\bf y}) = 2\pi k
\eqno(3.19)
$$
where $k$ is the winding number of the closed loop
${{\cal P}_x}{{{\cal P}'}_x^{-1}}$ around the point
${\bf y}^\star$.

The lattice extension of eq. (3.7) is not immediate;
it has been determined in [26] in quite great generality.
Here we present our version and make some general comments.
Let us choose as base point $B$ the point at infinity of
the positive $x$-axis, and let us associate to each point
${\bf x}$ the straight lattice path ${\cal C}_x$, parallel to
the $x$-axis from $B$ to ${\bf x}$ (see Fig. 4). Given any
two distinct points ${\bf x}$ and ${\bf y}$ and their
associated paths ${\cal C}_x$ and ${\cal C}_y$, it is
possible to show that
$$
\Theta_{{\cal C}_x}({\bf x},{\bf y}) -
\Theta_{{{\cal C}_y}}({\bf y},{\bf x}) =
\left\{\matrix{
 \pi \,{\rm sgn}(x_2-y_2) + \xi({\bf x},{\bf y}) ~~~~~
{\rm for}~~~x_2\not=y_2 \ \ , \cr
 \pi \,{\rm sgn}(x_1-y_1) + \xi({\bf x},{\bf y}) ~~~~~
{\rm for}~~~x_2=y_2 \ \ , \cr}\right.
\eqno(3.20)
$$
where the function $\xi({\bf x},{\bf y})$
is given in terms of the vector field ${\bf f}$ by [26]
$$
\xi({\bf x},{\bf y}) = -{1\over 2}
\left[f^1({\bf x},{\bf y})+f^2({\bf x},{\bf y})
+f^1({\bf x}+\hat{\bf 2},{\bf y}) + f^2({\bf x}
+\hat{\bf 1},{\bf y})\right]\ \ .
\eqno(3.21)
$$
Comparing eq. (3.20) with eq. (3.7), we notice that the extra
term $\xi({\bf x},{\bf y})$ appears in the right hand side. Such a
term is a genuine lattice feature arising from the fact that the
angles are measured from points of the dual lattice. This is
clearly seen in our explicit representation (3.13) where
$$
\xi({\bf x},{\bf y}) = \varphi({\bf x}+\hat{\bf 1}
+\hat{\bf 2},{\bf y}^\star,{\bf x}) \ \ .
\eqno(3.22)
$$
The right hand side represents the angle between
${\bf x}$ and ${\bf x}+\hat{\bf 1}+\hat{\bf 2}$ as seen
from ${\bf y}^\star$ which is equal to the angle between
the lines $({\bf x}\,{\bf y}^\star)$ and $({\bf x}^\star\,{\bf y})$,
as shown in Fig. 5.
A few remarks are in order. Eq. (3.14) implies that if ${\bf x}$
and ${\bf y}$ are very far apart from each other, $\xi({\bf x},{\bf y})$
is negligible and hence the left-hand side of eq. (3.20) simplifies
considerably. However this is not the case if ${\bf x}$ and
${\bf y}$ are close to each other. A possible way of removing the
$\xi$-term from eq. (3.20) even when ${\bf x}$ and ${\bf y}$ are not
far apart, is to embed the lattice $\Omega$ into another lattice
$\Lambda$ whose spacing $\varepsilon$ is taken to be much smaller
than one. Then, since $\Omega$ becomes a sublattice of $\Lambda$,
the quantities defined on $\Omega$ can be viewed as the restriction
to $\Omega$ of quantities defined on $\Lambda$. Under this assumption,
for all points ${\bf x},{\bf y}\in \Omega \subset \Lambda$ the vector
field ${\bf f}$ can be represented by
$$
f^i({\bf x},{\bf y})= \varphi({\bf x},{\bf y}',{\bf x}
+\varepsilon\,{\hat {\bf i}})
\eqno(3.23)
$$
where $\varphi({\bf x},{\bf y}',{\bf x}+\varepsilon\,
{\hat {\bf i}})$ is the angle under which the oriented link
of $\Lambda$ between ${\bf x}$ and ${\bf x}+\varepsilon\,
{\hat {\bf i}}$ is seen from the point ${\bf y}'=(y_1+
\varepsilon/2,
y_2+\varepsilon/2)$ belonging to the dual lattice
$\Lambda^\star$.
If we use the realization (3.23),
$\Theta_{{\cal C}_x}({\bf x},{\bf y})$ represents the
angle between $B$ and ${\bf x}$ measured from ${\bf y}'$
along the curve ${\cal C}_x$, while the function $\xi({\bf x},
{\bf y})$ becomes
$$
\xi({\bf x},{\bf y}) = \varphi({\bf x}+\varepsilon\,
{\hat {\bf 1}}+\varepsilon\,{\hat {\bf 2}},{\bf y}',{\bf x})
\ \ .
\eqno(3.24)
$$
In the limit $\varepsilon \to 0$, two distinct
points
${\bf x}$ and ${\bf y}$ are {\it always} far apart from each
other (from the standpoint of $\Lambda$) and the function $\xi({\bf x},
{\bf y})$ is {\it always} negligible.
Therefore eq. (3.20) simplifies to
$$
\Theta_{{\cal C}_x}({\bf x},{\bf y}) - \Theta_{{{\cal C}_y}}({\bf y},
{\bf x}) =
\left\{\matrix{
 \pi \,{\rm sgn}(x_2-y_2)  ~~~~~{\rm for}~~~x_2\not=y_2 \ \ , \cr
 \pi \,{\rm sgn}(x_1-y_1)  ~~~~~{\rm for}~~~x_2=y_2 \ \ ,
\cr}\right.
\eqno(3.25)
$$
which is the exact analogue of eq. (3.7) valid in the
continuum plane.
In this respect we remark that the choice of the curves
${\cal C}_x$ is in fact equivalent to fix for any ${\bf x}$
a cut $\gamma_x$ from ${\bf x}'\in \Lambda^\star$ to $-\infty$
along $x$-axis (see Fig. 6) in such a way that
$-\pi\le\Theta_{{\cal C}_x}<\pi$. This is the same fundamental
interval in which the continuum angle was defined.
Thus one can say that any point ${\bf x}\in \Omega$ is
characterized either by a curve ${\cal C}_x$ made of lattice
bonds or by a cut
$\gamma_x$ made of dual bonds.

We now use eq. (3.25) to establish an ordering relation among the
points of the lattice which later on will be essential to define
anyonic operators. First of all, let us introduce the notation
${\bf x}_{\cal C}$ to denote the point ${\bf x} \in \Omega$ with
its curve
${\cal C}_x$. Then, given two distinct points ${\bf x}$ and ${\bf y}$
we can posit
$$
{\bf x}_{\cal C}>{\bf y}_{\cal C} \Longleftrightarrow
\Theta_{{\cal C}_x}({\bf x},{\bf y}) -
\Theta_{{{\cal C}_y}}({\bf y},{\bf x}) =
\pi \ \ .
\eqno(3.26)
$$
Using eq. (3.25), we can rewrite this relation
more explicitly as follows
$$
{\bf x}_{\cal C}>{\bf y}_{\cal C}
\Longleftrightarrow
\left\{\matrix{
 x_2>y_2 \ \ , \cr
 x_2=y_2~,~x_1>y_1 \ \ . \cr}\right.
\eqno(3.27)
$$
Obviously, if $x_2<y_2$, or if $x_2=y_2$ and $x_1<y_1$,
then ${\bf x}_{\cal C}<{\bf y}_{\cal C}$. This definition is
unambiguous and endows the lattice $\Omega$ with an ordering
relation enjoying all the correct properties.
However this ordering is not unique. It crucially depends on
the choice of the curves
${\cal C}_x$. If we chose other types of curves, we could
clearly change the ordering relation. A more fundamental
change in the ordering can be obtained by modifying the definition
of the angle function. This is possible if we introduce a new
vector field $\tilde{\bf f}({\bf x},{\bf y})$ through
$$
\tilde f^i({\bf x},{\bf y})= - \epsilon^{ij}{\partial_j}
G({\bf x},{\bf y}) \ \ .
\eqno(3.28)
$$
This equation differs from that defining the old field
${\bf f}$ because the derivative $\tilde\partial_j$ has
been replaced by $\partial_j$ (cf eq. (3.11)). Moreover,
considering $\Omega$ as embedded into $\Lambda$, $\tilde
f^i({\bf x},{\bf y})$ is defined on the link between ${\bf x}$
and ${\bf x}-\varepsilon\hat{\bf i}$, in contrast with
$f^i({\bf x},{\bf y})$ which is defined on the link between
${\bf x}$ and ${\bf x}+\varepsilon\hat{\bf i}$. Keeping this
in mind, it is easy to show that
$$
\eqalign{
\tilde f^1({\bf x},{\bf y})-\tilde f^1({\bf x}-
\varepsilon{\hat{\bf 2}},{\bf y})-&
\tilde f^2({\bf x},{\bf y})+\tilde f^2({\bf x}-
\varepsilon{\hat{\bf 1}},{\bf y})
\cr
&=\epsilon^{ij}
\tilde\partial_i~\tilde f^j({\bf x},{\bf y})=
\tilde\partial_i \partial_i~G({\bf x},{\bf y}) =
2\pi\,\delta({\bf x},{\bf y}) \ \ ,}
\eqno(3.29)
$$
and hence
$$
\oint_{\tilde\Gamma_x}\tilde{\bf f}({\bf x},{\bf y})
=2\pi\,\delta({\bf x},{\bf y})
\eqno(3.30)
$$
where $\tilde\Gamma_x$ is the positively oriented
boundary of the elementary plaquette
$\tilde A_x$ of $\Lambda$ whose upper right corner
is the point ${\bf x}$ (see Fig. 7).
One possible representation of $\tilde{\bf f}$ is
the following
$$
\tilde f^i({\bf x},{\bf y})= \varphi({\bf x},
\tilde{\bf y},{\bf x}-\varepsilon\,{\hat {\bf i}})
\eqno(3.31)
$$
where $\varphi({\bf x},\tilde{\bf y},{\bf x}-
\varepsilon\,{\hat {\bf i}})$ is the angle under
which the oriented link of $\Lambda$
between ${\bf x}$ and ${\bf x}-\varepsilon\,{\hat {\bf i}}$
is seen from the point
$\tilde{\bf y}=(y_1-\varepsilon/2,
y_2-\varepsilon/2)\in \Lambda^\star$.
Then, given an arbitrary path ${\cal P}_x$ from
a base point $\tilde B$ to ${\bf x}$, we can define a new
angle function through
$$
\tilde\Theta_{{\cal P}_x}({\bf x},{\bf y}) =
\int_{{\cal P}_x} \tilde{\bf f}({\bf x},{\bf y}) \ \ .
\eqno(3.32)
$$
Analogously to eq. (3.17), if we use the
representation (3.31) we can interpret $
\tilde\Theta_{{\cal P}_x}({\bf x},{\bf y})$ as the
angle between $\tilde B$ and ${\bf x}$ measured from
$\tilde{\bf y}$ along the curve ${\cal P}_x$.
This angle function clearly satisfies
$$
\epsilon^{ij} \tilde\partial_i \tilde\partial_j~
\tilde\Theta_{{\cal P}_x}({\bf x},{\bf y})
=2\pi\,\delta({\bf x},{\bf y}) \ \ ,
\eqno(3.33)
$$
which is
another lattice version of eq. (3.5), and
$$
\tilde\Theta_{{\cal P}_x}({\bf x},{\bf y}) -
\tilde\Theta_{{{\cal P}'}_x}({\bf x},{\bf y}) =
\oint_{{{\cal P}_x}{{{\cal P}'}_x^{-1}}}
\tilde{\bf f}({\bf x},{\bf y}) = 2\pi k
\eqno(3.34)
$$
where $k$ is the winding number of the closed
loop ${{\cal P}_x}{{{\cal P}'}_x^{-1}}$ around the
point $\tilde{\bf y}\in \Lambda^\star$.

To write relations similar to eqs. (3.20) or (3.25)
for the function $\tilde\Theta$ we must again remove
all possible ambiguities; for instance this can be done
if we choose as base point $\tilde B$ the point at
infinity of the negative $x$-axis, and associate to
each point ${\bf x}$  the straight lattice path ${\cal D}_x$,
parallel to the $x$-axis from $\tilde B$ to ${\bf x}$ (see Fig. 8).
Then, with this choice and in the limit $\varepsilon\to 0$, one
can prove that for any two distinct points ${\bf x}$ and ${\bf y}$
$$
\tilde\Theta_{{\cal D}_x}({\bf x},{\bf y}) -
\tilde\Theta_{{{\cal D}_y}}({\bf y},{\bf x})
=\left\{\matrix{
 -\pi \,{\rm sgn}(x_2-y_2)  ~~~~~{\rm for}~~~x_2\not
=y_2 \ \ , \cr
 -\pi \,{\rm sgn}(x_1-y_1)  ~~~~~{\rm for}~~~x_2
=y_2 \ \ . \cr}\right.
\eqno(3.35)
$$
The choice of the curves ${\cal D}_x$ is therefore
equivalent to assign for any point ${\bf x}$ a cut
$\delta_x$ from $\tilde{\bf x}\in \Lambda^\star$ to $+\infty$
along the $x$-axis (see Fig. 9) in such a way that
$0\le {\tilde\Theta_{{\cal D}_x}}<2\pi$.

The relation (3.35) can be used to define a new
ordering among the points of the lattice. In fact, if
we denote by ${\bf x}_{\cal D}$ the point ${\bf x}$ with
its associated curve ${\cal D}_x$, in analogy with eq.
(3.26) we can posit
$$
{\bf x}_{\cal D}>{\bf y}_{\cal D} \Longleftrightarrow
\tilde\Theta_{{\cal D}_x}({\bf x},{\bf y}) -
\tilde\Theta_{{{\cal D}_y}}({\bf y},{\bf x}) =
\pi \ \ ,
\eqno(3.36)
$$
that is
$$
{\bf x}_{\cal D}>{\bf y}_{\cal D}
\Longleftrightarrow
\left\{\matrix{
 x_2<y_2 \ \ , \cr
 x_2=y_2~,~x_1<y_1 \ \ . \cr}\right.
\eqno(3.37)
$$
Comparing with eq. (3.27) we can easily realize
that the ordering defined by (3.37) is exactly the
opposite of the ordering induced by the curves ${\cal C}_x$.
Thus we have
$$
{\bf x}_{\cal C}>{\bf y}_{\cal C}
\Longleftrightarrow {\bf x}_{\cal D}<{\bf y}_{\cal D}
\ \ .
\eqno(3.38)
$$
For later convenience,
we now establish a direct relation between the ${\cal C}$-
and the ${\cal D}$-angles. A moment thought reveals that,
if ${\bf x}\not={\bf y}$
$$
\tilde\Theta_{{\cal D}_x}({\bf x},{\bf y}) -
\Theta_{{{\cal C}_x}}({\bf x},{\bf y})
=\left\{\matrix{
 -\pi \,{\rm sgn}(x_2-y_2)  ~~~~~{\rm for}~~~x_2\not
=y_2 \ \ , \cr
 -\pi \,{\rm sgn}(x_1-y_1)  ~~~~~{\rm for}~~~x_2=y_2
\ \ . \cr}\right.
\eqno(3.39)
$$
{}From eqs. (3.39) and (3.35) it follows that
$$
\tilde\Theta_{{{\cal D}_x}}({\bf x},{\bf y})
-\Theta_{{{\cal C}_y}}({\bf y},{\bf x}) = 0
\eqno(3.40)
$$
for {\it all} ${\bf x}\not={\bf y}$.

Actually, eq. (3.40) is formally correct also
for ${\bf x}={\bf y}$. Indeed,
$\tilde\Theta_{{{\cal D}_x}}({\bf x},{\bf x})=
-{3\pi\over 4}$ and $\Theta_{{{\cal C}_x}}({\bf x},{\bf x})=
-{3\pi\over 4}$ so that
$$\tilde\Theta_{{\cal D}_x}({\bf x},{\bf x}) -
\Theta_{{{\cal C}_x}}({\bf x},{\bf x}) = 0 \ \ .
\eqno(3.41)
$$

We can summarize our findings by saying that the set of the
lattice points is doubled into points of the ${\cal C}$-type
({\it i.e.} ${\bf x}_{\cal C}$ for ${\bf x}\in \Omega$) and into
points of the ${\cal D}$-type ({\it i.e.} ${\bf x}_{\cal D}$ for
${\bf x}\in \Omega$) which are ordered among themselves according
to eqs. (3.27) and (3.37) respectively. As we will show in the
next section, this doubling of the lattice points
is crucial in the construction of the quantum group structure.

After this discussion on the lattice angle function and on
the ordering relations induced by it, we are finally ready
to introduce anyonic operators on $\Omega$. These are defined
by means of a (generalized) Jordan-Wigner transformation from
the fermionic operators $c_1({\bf x}), c_2({\bf x})$ and their
adjoints which were considered in the previous section. Roughly
speaking, the Jordan-Wigner transformation amounts to simply
stick a disorder operator to the fermions $c_i({\bf x})$ [21,28,29].
Such a disorder operator can be written as the exponential of the
local fermion density $c^\dagger_i({\bf x})c_i({\bf x})$ summed
over all lattice points and weighted with the angle function.
However, since we have defined two lattice angle functions, we
should expect two types of disorder operators, and hence two
types of lattice anyons (type ${\cal C}$ and type ${\cal D}$).

Let us first define the anyons of type ${\cal C}$
according to
$$
a_i({\bf x}_{\cal C}) = K_i({\bf x}_{\cal C})
\,c_i({\bf x}) \ \ ,
\eqno(3.42{\rm a})
$$
and
$$
a^\dagger_i({\bf x}_{\cal C}) =
c^\dagger_i({\bf x})\,
K^\dagger_i({\bf x}_{\cal C})
\eqno(3.42{\rm b})
$$
where
$$
\eqalign{
K_i({\bf x}_{\cal C}) &= {\rm e}^{{\rm i}
\nu\sum\limits_{{\bf y}\in \Omega}
\Theta_{{{\cal C}_x}}({\bf x},{\bf y})\,
c^\dagger_i({\bf y})c_i({\bf y})} \ \ , \cr
K^\dagger_i({\bf x}_{\cal C}) &=
{\rm e}^{-{\rm i}\nu\sum\limits_{{\bf y}\in \Omega}
\Theta_{{{\cal C}_x}}({\bf x},{\bf y})\,
c^\dagger_i({\bf y})c_i({\bf y})}=
K^{-1}_i({\bf x}_{\cal C}) \cr}
\eqno(3.43)
$$
are disorder operators [21,28]
\footnote{${}^3$}{{Notice that
when plugging the expressions (3.43) into eqs.
(3.42), the term with ${\bf y}={\bf x}$ in the sum of
the exponent does {\it not} contribute
since $\Big[c_i({\bf x})\Big]^2=
\left[c_i^\dagger({\bf x})\right]^2=0$; we remark
however that $\Theta_{{{\cal C}_x}}({\bf x},{\bf y})$ is
defined also for ${\bf y}={\bf x}$.}}. In these formulas
$i=1,2$ for ``spin-up'' and ``spin-down'' respectively,
and
$\nu$ is a real parameter which, as we will see, represents
the statistics (our conventions are such that for $\nu=0$ we
have fermionic statistics, whereas for $\nu=1$ we have bosonic
statistics).

Using the canonical commutation relations of the fermionic
operators (cf eq. (2.12)), it is very easy to show that
$$
\eqalign{
K_i({\bf x}_{\cal C})~c_i({\bf y}) &=
{\rm e}^{-{\rm i}\nu\Theta_{{{\cal C}_x}}({\bf x},{\bf y})}
{}~c_i({\bf y})~K_i({\bf x}_{\cal C}) \ \ , \cr
K_i({\bf x}_{\cal C})~c^\dagger_i({\bf y}) &=
{\rm e}^{{\rm i}\nu\Theta_{{{\cal C}_x}}
({\bf x},{\bf y})}~c^\dagger_i({\bf y})~K_i({\bf x}_{\cal C})\ \ , \cr
K_i({\bf x}_{\cal C})~K_i({\bf y}_{\cal C})&=
K_i({\bf y}_{\cal C})~K_i({\bf x}_{\cal C})\ \ ,}
\eqno(3.44)
$$
for all ${\bf x}$ and ${\bf y}$.

To see that the operators $a_i({\bf x}_{\cal C})$
and $a^\dagger_i({\bf x}_{\cal C})$  are indeed anyons
of statistics $\nu$, let us compute their ``commutation''
relations. Let us also simplify the notations and introduce
the symbol ${\bf x}>{\bf y}$ to mean
${\bf x}_{\cal C}>{\bf y}_{\cal C}$ (and hence also
${\bf x}_{\cal D}<{\bf y}_{\cal D}$). Thus, for ${\bf x}>{\bf y}$
we have
$$
\eqalign{
a_i({\bf x}_{\cal C})~a_i({\bf y}_{\cal C})&
=K_i({\bf x}_{\cal C})
\,c_i({\bf x})~K_i({\bf y}_{\cal C})
\,c_i({\bf y})\cr
&=-{\rm e}^{-{\rm i}\nu\left[
\Theta_{{{\cal C}_x}}({\bf x},{\bf y}) -
\Theta_{{{\cal C}_y}}({\bf y},{\bf x})\right]}~
K_i({\bf y}_{\cal C})\,c_i({\bf y})~
K_i({\bf x}_{\cal C})\,c_i({\bf x})\cr
&=-{\rm e}^{-{\rm i}\nu\pi}~a_i({\bf y}_{\cal C})~
a_i({\bf x}_{\cal C})\ \ ,}
\eqno(3.45)
$$
where to derive the last line we used eqs. (3.44) and (3.26).
This formula can be written also as
$$
a_i({\bf x}_{\cal C})~a_i({\bf y}_{\cal C})+q^{-1}~
a_i({\bf y}_{\cal C})~a_i({\bf x}_{\cal C})=0
\eqno(3.46{\rm a})
$$
where
$$
q={\rm e}^{{\rm i}\nu\pi} \ \ .
\eqno(3.47)
$$
For $\nu=0~({\rm mod}\,2)$ eq. (3.45a) is an
anticommutator signaling fermionic statistics,
whilst for $\nu=1~({\rm mod}\,2)$ it is a commutator
signaling bosonic statistics.
The embedding of the lattice $\Omega$ into $\Lambda$ and the
limiting procedure
$\varepsilon \to 0$ which allowed us to eliminate the $\xi$-term
from eq. (3.20), are essential to get a generalized commutator
with a constant $q$-factor; otherwise one would obtain eq. (3.46a)
with $q$ depending on the distance ${\bf x}-{\bf y}$. For an
alternative procedure to remove this dependence and the ${\xi}$-term
see [26].
With similar calculations we can derive also the following
generalized commutation relations
$$
\eqalignno{
a^\dagger_i({\bf x}_{\cal C})~
a^\dagger_i({\bf y}_{\cal C})&+q^{-1}~
a^\dagger_i({\bf y}_{\cal C})~
a^\dagger_i({\bf x}_{\cal C})=0 \ \ ,
&(3.46{\rm b})\cr
a_i({\bf x}_{\cal C})~a^\dagger_i({\bf y}_{\cal C})
&+q~a^\dagger_i({\bf y}_{\cal C})~a_i({\bf x}_{\cal C})=0
\ \ , &(3.46{\rm c})\cr
a^\dagger_i({\bf x}_{\cal C})~a_i({\bf y}_{\cal C})
&+q~a_i({\bf y}_{\cal C})~a^\dagger_i({\bf x}_{\cal C})=0
\ \ , &(3.46{\rm d})\cr}
$$
for all ${\bf x}>{\bf y}$. We notice that eqs. (3.46b)
and (3.46d) are the hermitian conjugate of eqs. (3.46a)
and (3.46c) respectively, since $q^\star=q^{-1}$. For
completeness we recall that
$$
\Big[a_i({\bf x}_{\cal C})\Big]^2=\left[
a^\dagger_i({\bf x}_{\cal C})\right]^2=0\ \ ,
\eqno(3.48{\rm a})
$$
and
$$
\eqalign{
\Big\{
a_1({\bf x}_{\cal C})~,~a_2({\bf y}_{\cal C})\Big\}&
=\left\{a^\dagger_1({\bf x}_{\cal C})~,
{}~a^\dagger_2({\bf y}_{\cal C})\right\}\cr
=\left\{a^\dagger_1({\bf x}_{\cal C})~,
{}~a_2({\bf y}_{\cal C})\right\}&=\left\{a_1({\bf x}_{\cal C})~,
{}~a^\dagger_2({\bf y}_{\cal C})\right\}=0}
\eqno(3.48{\rm b})
$$
for all ${\bf x}$ and ${\bf y}$ and for any value of $\nu$.
Eq. (3.48a) enforces the Pauli exclusion principle;
thus for $\nu=1~({\rm mod}\,2)$ the oscillator
$a_i({\bf x}_{\cal C})$ represents a boson with a
hard core. The (anti)commutation relations of
$a_i$ and $a^\dagger_i$ in the same point deserve
particular attention. In fact, using the definitions
(3.42) and (3.43) and the relations (3.44), we have
$$
\eqalign{
a_i({\bf x}_{\cal C})~a^\dagger_i({\bf x}_{\cal C})
&=K_i({\bf x}_{\cal C})\,
c_i({\bf x})~c^\dagger_i({\bf x})\,
K^\dagger_i({\bf x}_{\cal C}) \cr
&=K_i({\bf x}_{\cal C})\,
\left(-c^\dagger_i({\bf x})~c_i({\bf x})+1\right)\,
K^\dagger_i({\bf x}_{\cal C})\cr
&=-{\rm e}^{{\rm i}\nu\Theta_{{{\cal C}_x}}({\bf x},{\bf x})}
{}~c^\dagger_i({\bf x})\,K_i({\bf x}_{\cal C})~~
{\rm e}^{-{\rm i}\nu\Theta_{{{\cal C}_x}}({\bf x},{\bf x})}~
K^\dagger_i({\bf x}_{\cal C})\,c_i({\bf x})~+~1\cr
&=-a^\dagger_i({\bf x}_{\cal C})~
a_i({\bf x}_{\cal C})~+~1 \ \ .}
\eqno(3.49)
$$
In conclusion we find
$$
a_i({\bf x}_{\cal C})~a^\dagger_i({\bf x}_{\cal C})
+a^\dagger_i({\bf x}_{\cal C})~a_i({\bf x}_{\cal C})=1
\eqno(3.50)
$$
without any phase factor. Therefore,
contrarily to several statements in the
literature [21,25,26,28], we see that the fermionic
based operators $a_i$ and $a^\dagger_i$ with anyonic
statistics $\nu$
obey {\it standard} anticommutation relations at the
same point
\footnote{${}^4$}{{We remark that one consistently obtains
the same result (3.50) also if one defines $K_i({\bf x}_{\cal C})$
by excluding the point ${\bf y}={\bf x}$ from the sum at
exponent in eq. (3.43); in such a case then, $K_i({\bf x}_{\cal C})$
would commute with both $a_i({\bf x}_{\cal C})$ and
$a^\dagger_i({\bf x}_{\cal C})$.}}. Furthermore, since
anyons carry a representation of the braid group (see
for instance [4]), when we exchange two of them it is essential
to specify the orientation of the exchanging trajectories
({\it i.e.} their braidings), and on the lattice this can be
done unambiguously only by exploiting the ordering induced by
the lattice angle function. In fact the exchange of two anyons
located in ${\bf x}$ and ${\bf y}$ can be realized by a half-circle
rotation of ${\bf y}$ around ${\bf x}$ followed by a rigid translation.
The orientation of such rotation can be uniquely defined by requiring
that ${\bf y}$ does not cross the cut ${\gamma}_x$. One can easily
get convinced that the rotation is counterclockwise for
${\bf x}>{\bf y}$
and clockwise for ${\bf x}<{\bf y}$ anyons (see Fig. 10).
This aspect has not been sufficiently emphasized in the previous
literature on this subject [21,25,26,28], but after all, also in
the continuum theory one has to specify if the anyons are
exchanged counterclockwise or clockwise to define
unambiguously their statistics !

Let us now define the anyon operators of type ${\cal D}$.
They are given by
$$
a_i({\bf x}_{\cal D}) = K_i({\bf x}_{\cal D})
\,c_i({\bf x}) \ \ ,
\eqno(3.51{\rm a})
$$
and
$$
a^\dagger_i({\bf x}_{\cal D}) = c^\dagger_i({\bf x})\,
K^\dagger_i({\bf x}_{\cal D})
\eqno(3.51{\rm b})
$$
where the type ${\cal D}$ disorder operators are
$$
\eqalign{
K_i({\bf x}_{\cal D}) &=
{\rm e}^{{\rm i}\nu\sum\limits_{{\bf y}\in \Omega}
\tilde\Theta_{{{\cal D}_x}}({\bf x},{\bf y})\,
c^\dagger_i({\bf y})c_i({\bf y})} \ \ , \cr
K^\dagger_i({\bf x}_{\cal D}) &=
{\rm e}^{-{\rm i}\nu\sum\limits_{{\bf y}\in \Omega}
\tilde\Theta_{{{\cal D}_x}}({\bf x},{\bf y})
\,c^\dagger_i({\bf y})c_i({\bf y})}=
K^{-1}_i({\bf x}_{\cal D}) \ \ .\cr}
\eqno(3.52)
$$
Clearly these disorder operators obey the same
relations as in eq. (3.44) with
${\bf x}_{\cal C}$ and $\Theta_{{{\cal C}_x}}$ replaced
by ${\bf x}_{\cal D}$ and $\tilde\Theta_{{{\cal D}_x}}$
respectively. With manipulations similar to those that led to
eqs. (3.46),
it is easy to show that
$$
\eqalignno{
a_i({\bf x}_{\cal D})~a_i({\bf y}_{\cal D})
&+q~a_i({\bf y}_{\cal D})~a_i({\bf x}_{\cal D})=0 \ \ ,
&(3.53{\rm a})\cr
a^\dagger_i({\bf x}_{\cal D})~a^\dagger_i({\bf y}_{\cal D})
&+q~a^\dagger_i({\bf y}_{\cal D})~a^\dagger_i({\bf x}_{\cal D})=
0 \ \ , &(3.53{\rm b})\cr
a_i({\bf x}_{\cal D})~a^\dagger_i({\bf y}_{\cal D})
&+q^{-1}~a^\dagger_i({\bf y}_{\cal D})~a_i({\bf x}_{\cal D})=0 \ \ ,
&(3.53{\rm c})\cr
a^\dagger_i({\bf x}_{\cal D})~a_i({\bf y}_{\cal D})
&+q^{-1}~a_i({\bf y}_{\cal D})~a^\dagger_i({\bf x}_{\cal D})=0 \ \ ,
&(3.53{\rm d})\cr}
$$
for all ${\bf x}>{\bf y}$. Notice that ${\bf x}>{\bf y}$ means
${\bf x}_{\cal C}>{\bf y}_{\cal C}$ and hence ${\bf x}_{\cal D}<
{\bf y}_{\cal D}$ (cf eq. (3.38)). Furthermore
$$
a_i({\bf x}_{\cal D})~a^\dagger_i({\bf x}_{\cal D})
+a^\dagger_i({\bf x}_{\cal D})~a_i({\bf x}_{\cal D})=1 \ \ ,
\eqno(3.54)
$$
{\it i.e.} they satisfy standard anticommutation relations at
the same point.
Eqs. (3.53) must be interpreted by saying that $a_i({\bf x}_{\cal D})$
and
$a^\dagger_i({\bf x}_{\cal D})$ are again anyons of statistics $\nu$.
However we see that these generalized commutation relations differ from
the corresponding ones for the type ${\cal C}$ operators (cf eqs.
(3.46)) because $q$ has been replaced by $q^{-1}$.
This should not come as a
surprise because the ${\cal C}$ and the ${\cal D}$ orderings are
inverse to one another. More precisely, one can say that the ${\cal D}$
 ordering can be obtained from the ${\cal C}$ ordering with a
parity transformation which, as well known, changes the braiding
phase $q$ into $q^{-1}$ (see for instance [4]). Indeed Fig. 11 clearly
shows that the opposite orientation of the cuts $\delta_x$ with
respect to the cuts $\gamma_x$ reverses also the orientation of the
exchanging trajectories.

It is now interesting to establish a direct relation between the
${\cal C}$ and the ${\cal D}$ operators. Repeatedly using their
definitions and eq. (3.44) with its analogue for the ${\cal D}$
operators, we find
$$
\eqalign{
a_i({\bf x}_{\cal D})~a_i({\bf y}_{\cal C}) &=
K_i({\bf x}_{\cal D})\,c_i({\bf x})~K_i({\bf y}_{\cal C})
\,c_i({\bf y})\cr
&=-{\rm e}^{-{\rm i}\nu\left[\tilde\Theta_{{{\cal D}_x}}({\bf x},
{\bf y})-\Theta_{{{\cal C}_y}}({\bf y},{\bf x})\right]}~~
a_i({\bf y}_{\cal C})~a_i({\bf x}_{\cal D})\ \ .}
\eqno(3.55)
$$
The exponent in the last line actually vanishes
for all ${\bf x}$ and ${\bf y}$ because of eq. (3.40).
Thus we get
$$
\Big\{a_i({\bf x}_{\cal D})~,~a_i({\bf y}_{\cal C})\Big\}=
0~~~~~~~~~~~\forall ~{\bf x},\,{\bf y}\ \ .
\eqno(3.56{\rm a})
$$
Similarly we have
$$
\Big\{a_i({\bf x}_{\cal D})~,~
a^\dagger_i({\bf y}_{\cal C})\Big\}=0~~~~~~~~~~~~\forall
{}~{\bf x}\not={\bf y}\ \ .
\eqno(3.56{\rm b})
$$
By taking the hermitian conjugate of these expressions,
we obtain
$$
\eqalignno{
\Big\{a^\dagger_i({\bf x}_{\cal D})~,
{}~a^\dagger_i({\bf y}_{\cal C})\Big\}&=0~~~~~~~~~~~~
\forall ~{\bf x},\,{\bf y}\ \ ,
&(3.56{\rm c})\cr
\Big\{a^\dagger_i({\bf x}_{\cal D})~,~
a_i({\bf y}_{\cal C})\Big\}&=0~~~~~~~~~~~~
\forall ~{\bf x}\not={\bf y}\ \ .
&(3.56{\rm d})}
$$

Things are not so simple in the anticommutation
relation of $a_i({\bf x}_{\cal D})$ and
$a^\dagger_i({\bf x}_{\cal C})$, {\it i.e.} at the
same point. Indeed we have
$$
\eqalign{
a_i({\bf x}_{\cal D})~a^\dagger_i({\bf x}_{\cal C})&=
K_i({\bf x}_{\cal D})\,c_i({\bf x})~c^\dagger_i({\bf x})
\,K^\dagger_i({\bf x}_{\cal C})\cr
&=K_i({\bf x}_{\cal D})\,\left(-c^\dagger_i({\bf x})~
c_i({\bf x})~+~1\right)\,
K^\dagger_i({\bf x}_{\cal C})\cr
&=-{\rm e}^{{\rm i}\nu\left[
\tilde\Theta_{{{\cal D}_x}}({\bf x},{\bf x})-
\Theta_{{{\cal C}_x}}({\bf x},{\bf x})\right]}
{}~~a^\dagger_i({\bf x}_{\cal C})~
a_i({\bf x}_{\cal D})
{}~+~K_i({\bf x}_{\cal D})\,
K^\dagger_i({\bf x}_{\cal C})\cr
&= -~a^\dagger_i({\bf x}_{\cal C})~
a_i({\bf x}_{\cal D})
{}~+~K_i({\bf x}_{\cal D})\,
K^\dagger_i({\bf x}_{\cal C})}
\eqno(3.57)
$$
where in the final step we made use of eq. (3.41).
Inserting the explicit expressions of the disorder
operators, we can simplify the last term and get
$$
\eqalign{
K_i({\bf x}_{\cal D})\,K^\dagger_i({\bf x}_{\cal C})&=
{\rm e}^{{\rm i}\nu\sum\limits_{{\bf y}\in \Omega}
\Big[\tilde\Theta_{{{\cal D}_x}}({\bf x},{\bf y})-
\Theta_{{{\cal C}_x}}({\bf x},{\bf y})\Big]
c^\dagger_i({\bf y})\,c_i({\bf y})} \cr
&={\rm e}^{-{\rm i}\nu\pi
\Big[\sum\limits_{{\bf y}<{\bf x}}-
\sum\limits_{{\bf y}>{\bf x}}\Big]
c^\dagger_i({\bf y})\,c_i({\bf y})} }
\eqno(3.58)
$$
where eqs. (3.39) and (3.41) have been
taken into account. Combining the last two
equations we obtain
$$
\Big\{a_i({\bf x}_{\cal D})~,
{}~a^\dagger_i({\bf x}_{\cal C})\Big\}=
q^{-\Big[\sum\limits_{{\bf y}<{\bf x}}-
\sum\limits_{{\bf y}>{\bf x}}\Big]
c^\dagger_i({\bf y})\,c_i({\bf y})} \ \ ;
\eqno(3.59)
$$
finally, taking its hermitian
conjugate yields
$$
\Big\{a_i({\bf x}_{\cal C})~,
{}~a^\dagger_i({\bf x}_{\cal D})\Big\}=
q^{\Big[\sum\limits_{{\bf y}<{\bf x}}-
\sum\limits_{{\bf y}>{\bf x}}\Big]
c^\dagger_i({\bf y})\,c_i({\bf y})} \ \ .
\eqno(3.60)
$$

This concludes our discussion of anyon oscillators on
the lattice; in the next section we will use them to
realize the quantum group $SU(2)_q$ with the Schwinger
construction, and will find that in order to close correctly
the quantum algebra it is essential the employ both the type
${\cal C}$ and the type ${\cal D}$ operators. Therefore these
two types of lattice anyons which, as we have mentioned, are
related to one another by a parity transformation, will
find an algebraic application in a quite natural way.

\vfill
\eject
\centerline{\bf 4. The Schwinger Construction with
 Anyons and}
\centerline{\bf the Quantum Group $SU(2)_q$}
\vskip 1cm
In this section we are going to show that the Schwinger
construction of $SU(2)$ which was previously discussed
using bosons and fermions, can be generalized to anyons of
statistics $\nu$ in a quite direct though non-trivial way.
However in the case of anyons, the Schwinger construction will
not realize the ordinary group $SU(2)$ but rather its quantum
deformation $SU(2)_q$ with $q=\exp({\rm i}\pi\nu)$.

In analogy with the bosonic formula (2.2), or even better with
the fermionic ones (2.13) and (2.15), we start by introducing
the local step-up operator
$$
J^+({\bf x}) = a^\dagger_1({\bf x}_{\cal C})\,
a_2({\bf x}_{\cal C}) \ \ ,
\eqno(4.1)
$$
which, as an immediate consequence of eqs. (3.46),
satisfies
$$
J^+({\bf y})~J^+({\bf x})=q^2~J^+({\bf x})~
J^+({\bf y})
\eqno(4.2)
$$
for ${\bf x}>{\bf y}$. This formula is quite important
for several reasons. First of all, it shows that these
operators have braiding properties just like their constituent
factors. Therefore, in discussing their generalized commutation
relations it is crucial to specify the ordering of the points;
indeed the explicit $q$-factor in the right hand side changes if
we change the braiding orientation, {\it i.e.} the ordering of
${\bf x}$ and ${\bf y}$. Secondly, and also in view of the last
observation, it should be clear that if we used type ${\cal D}$ anyons
instead of type ${\cal C}$ ones, we would make a parity transformation
reversing the ordering of the points and thus we would change $q$ into
$q^{-1}$. In fact the operator
$$
\tilde{J}^+({\bf x}) = a^\dagger_1({\bf x}_{\cal D})\,
a_2({\bf x}_{\cal D})
\eqno(4.3)
$$
satisfies
$$
\tilde{J}^+({\bf y})~\tilde{J}^+({\bf x})=
q^{-2}~\tilde{J}^+({\bf x})
{}~\tilde{J}^+({\bf y})
\eqno(4.4)
$$
for ${\bf x}>{\bf y}$, as one can check using the
generalized commutation relations (3.53). Finally, we remark that
the behaviour of $J^+({\bf x})$ exhibited in eq. (4.2) is the same
as the one of the local densities of quantum group generators. By
this we mean that if $J_q^+=\sum\limits_{{\bf x}}J_q^+({\bf x})$ is
a generator of the quantum group $SU(2)_q$ obtained by repeated use
of comultiplication starting from the local operators $j^+({\bf x})$
and $j^0({\bf x})$, then $J_q^+({\bf y})\,J_q^+({\bf x})=q^2\,
J_q^+({\bf x})\,J_q^+({\bf y})$ for ${\bf x}>{\bf y}$ (see
for example
[9,10]). This is a significant hint that actually the
step-up
operator (4.1) can be somehow considered as the local density
of a quantum group
generator, {\it i.e.}
$$
J^+({\bf x}) \simeq J_q^+({\bf x}) \ \ .
\eqno(4.5)
$$
Similarly from eq. (4.4) one may say that
$$
\tilde{J}^+({\bf x}) \simeq J_{q^{-1}}^+({\bf x}) \ \ ,
\eqno(4.6)
$$
$J_{q^{-1}}^+$ being the generator of the quantum group
$SU(2)_{q^{-1}}$.

If we pursue this conjecture further on, we should expect that
the step-down operator $J^-({\bf x})$ be related to the step-up
operator $J^+({\bf x})$ like the local densities of the quantum
group generators, namely like
$$
J_q^-({\bf x})=\left[J_{q^\star}^+({\bf x})\right]^\dagger
\eqno(4.7)
$$
(see for example [9] and eq. (2.22)). In our case $q^\star=q^{-1}$,
and thus we are led to posit
$$
J^-({\bf x}) = \left[\tilde{J}^+({\bf x})\right]^\dagger
=a^\dagger_2({\bf x}_{\cal D})\,a_1({\bf x}_{\cal D}) \ \ .
\eqno(4.8)
$$
Using eqs. (3.53), one easily proves that
$$
J^-({\bf y})~J^-({\bf x})=q^{-2}~J^-({\bf x}) ~
J^-({\bf y})
\eqno(4.9)
$$
for ${\bf x}>{\bf y}$, as expected.

Inspired by the ordinary Schwinger construction
(cf eqs. (2.2) and (2.13)), we may define the Cartan generator
$J_q^0({\bf x})$ according to
$$
J^0({\bf x})={1\over2}\left(a^\dagger_1({\bf x}_{\cal C})\,
a_1({\bf x}_{\cal C})-
a^\dagger_2({\bf x}_{\cal C})\,a_2({\bf x}_{\cal C})
\right)\ \ .
\eqno(4.10)
$$
It is interesting to realize that in this expression we
could have used the anyon oscillators of type ${\cal D}$
without any change; in fact one may check that the disorder
operators cancel out yielding
$$
a^\dagger_i({\bf x}_{\cal D})\,a_i({\bf x}_{\cal D})=
a^\dagger_i({\bf x}_{\cal C})\,a_i({\bf x}_{\cal C}) =
c^\dagger_i({\bf x})\,c_i({\bf x})
\ \ .
\eqno(4.11)
$$
Thus, $J^0({\bf x})$ does not depend on $q$ and is the same
as in the fermionic
realization (see eq. (2.13)).
In conclusion, one may say that the anyonic generalization of
the Schwinger costruction leads to consider the following three
local operators
$$
\eqalign{
J^+({\bf x}) &= a^\dagger_1({\bf x}_{\cal C})\,
a_2({\bf x}_{\cal C}) \ \ ,\cr
J^0({\bf x})&={1\over2}\left(a^\dagger_1({\bf x}_{\cal C})\,
a_1({\bf x}_{\cal C})-
a^\dagger_2({\bf x}_{\cal C})\,a_2({\bf x}_{\cal C})\right)\cr
&={1\over2}\left(a^\dagger_1({\bf x}_{\cal D})\,
a_1({\bf x}_{\cal D})-
a^\dagger_2({\bf x}_{\cal D})\,
a_2({\bf x}_{\cal D})\right)\ \ ,\cr
J^-({\bf x}) &= a^\dagger_2({\bf x}_{\cal D})\,
a_1({\bf x}_{\cal D})\ \ . }
\eqno(4.12)
$$
What remains to be discussed is what algebra, if any,
such operators close.
We have already noticed and conjectured a sort of relation
between these operators and the local densities of quantum
group generators; hereinafter we are going to prove that this
conjecture is correct.

With a straightforward application of eqs. (3.46), (3.53) and (3.56),
we can easily
check that
$$
\eqalign{
\left[ J^0({\bf x})~,~J^{\pm}({\bf y})\right] &=\pm
J^{\pm}({\bf x}) ~
\delta({\bf x},{\bf y}) \ \ , \cr
\left[ J^+({\bf x})~,~J^-({\bf y})\right]&=0~~~~~~~~~~~~~~
\forall\,{\bf x}\not={\bf y} \ \ .}
\eqno(4.13)
$$
The commutation relation of $J^+({\bf x})$ with $J^-({\bf x})$
({\it i.e.} at the same point) is slightly more complicated because
it envolves the anticommutators of anyons of type ${\cal C}$ with
anyons of type ${\cal D}$. However, using eqs. (3.59) and (3.60),
it is not difficult to show that
$$
\eqalign{
\left[ J^+({\bf x})~,~J^-({\bf x})\right] =&
\,q^{\Big[\sum\limits_{{\bf y}<{\bf x}}-
\sum\limits_{{\bf y}>{\bf x}}\Big]c^\dagger_2({\bf y})\,
c_2({\bf y})}~a^\dagger_1({\bf x}_{\cal C})\,
a_1({\bf x}_{\cal D})\cr
&-q^{-\Big[\sum\limits_{{\bf y}<{\bf x}}-
\sum\limits_{{\bf y}>{\bf x}}\Big]c^\dagger_1({\bf y})\,
c_1({\bf y})}~a^\dagger_2({\bf x}_{\cal D})\,a_2({\bf x}_{\cal C})
\ \ .}
\eqno(4.14{\rm a})
$$
If we insert the explicit definition of anyon oscillators
in the right hand side and then use eq. (3.39), this commutator
can be rewritten in a more useful form as follows
$$
\left[ J^+({\bf x})~,~J^-({\bf x})\right] =
\prod_{{\bf y}<{\bf x}}q^{-2J^0({\bf y})}~2J^0({\bf x})~
\prod_{{\bf z}>{\bf x}}q^{2J^0({\bf z})}\ \ .
\eqno(4.14{\rm b})
$$

After these preliminaries we are now in the position of
defining the global generators. These are given by
$$
\eqalign{
J^\pm &=\sum_{{\bf x}\in \Omega}J^\pm({\bf x}) \ \ ,
\cr
J^0 &=\sum_{{\bf x}\in \Omega}J^0({\bf x}) \ \ ,}
\eqno(4.15)
$$
and close the $SU(2)_q$ algebra. In fact, from eqs.
(4.13) and (4.14) one easily obtains
$$
\eqalignno{
\left[J^0~,~J^\pm\right]&=\pm J^\pm \ \ ,
&(4.16{\rm a})\cr
\left[J^+~,~J^-\right]&=\sum_{{\bf x}\in \Omega}
\left(\prod_{{\bf y}<{\bf x}}q^{-2J^0({\bf y})}~
2J^0({\bf x})~\prod_{{\bf z}>{\bf x}}
q^{2J^0({\bf z})}\right)
\ \ . &(4.16{\rm b})}
$$
These are precisely the defining commutation relations of
$SU(2)_q$ when $J^0({\bf x})$ is in the spin 0 or spin 1/2
representation for any ${\bf x}$. In the literature on quantum
groups one usually finds a different expression for the last
commutator, namely
$$
\left[J^+~,~J^-\right]={{q^{2J^0}-q^{-2J^0}}\over{q-q^{-1}}}
\eqno(4.17)
$$
(see for instance [7-9]). Despite the appearance, there is
actually no difference between eqs. (4.16b) and (4.17); in
fact one can prove that in our case
$$
\sum_{{\bf x}\in \Omega}\left(\prod_{{\bf y}<{\bf x}}
q^{-2J^0({\bf y})}~2J^0({\bf x})~\prod_{{\bf z}>{\bf x}}
q^{2J^0({\bf z})}\right) = {{q^{2J^0}-q^{-2J^0}}
\over{q-q^{-1}}} \ \ .
\eqno(4.18)
$$
To prove this equality we use the method of complete
induction.
First of all one has to realize that our operator $J^0({\bf x})$
admits only the eigenvalues 0 and $\pm 1/2$ for any ${\bf x}$.
This fact is a direct consequence of the Pauli exclusion principle
for anyon operators, which, in this respect, behave like ordinary
fermions
(cf eq. (3.48a)). Therefore, the following identity holds
$$
2\,J^0({\bf x}) = {{q^{2J^0({\bf x})}-q^{-2J^0({\bf x})}}
\over{q-q^{-1}}}
\eqno(4.19)
$$
for any ${\bf x}$.
If the lattice $\Omega$ has only one site, say the point ${\bf a}$,
clearly $J^0=J^0({\bf a})$ and thus the equality (4.18) is true by
virtue of the identity (4.19).

Let us now assume that eq. (4.18) is correct for a lattice
$\Omega$ of $N$ sites and then prove that it remains true when
an extra point, say ${\bf b}$, is added. We suppose that
$$
{\bf b} > {\bf x}_i
$$
for $i=1,...,N$, so that eq. (4.18) for the lattice with
$N+1$ sites becomes
$$
\eqalign{
\sum_{i=1}^N&\left[\prod_{j<i}q^{-2J^0({\bf x}_j)}~
2J^0({\bf x}_i)~\prod_{k>i}q^{2J^0({\bf x}_k)}\right]~
q^{2J^0({\bf b})}
+~q^{-2J^0}~2J^0({\bf b})\cr
&~~~~~~~~~~~~~~~~~~~={1\over{q-q^{-1}}}\left(q^{2J^0}\,
q^{2J^0({\bf b})}-q^{-2J^0}\,q^{-2J^0({\bf b})}\right)\ \ ,}
\eqno(4.20)
$$
where $J^0=\sum\limits_{i=1}^NJ^0({\bf x}_i)$. Eq. (4.20)
is easily proved using eq. (4.18) to replace the term in square
brackets in the left hand side with
$$
{{q^{2J^0}-q^{-2J^0}}\over{q-q^{-1}}} \ \ ,
$$
so that one is left with
$$
q^{-2J^0}~2J^0({\bf b})={{q^{-2J^0}}\over{q-q^{-1}}}
\left(q^{2J^0({\bf b})}-q^{-2J^0({\bf b})}\right) \ \ ,
$$
which is true because of the identity (4.19).
The same result can be obtained also when ${\bf b}<{\bf x}_i$.
This concludes our proof of eq. (4.18).

Therefore we have explicitly shown that the operators $J^\pm$
and $J^0$ built out of anyon oscillators by means of the generalized
Schwinger construction do close the algebra of $SU(2)_q$ where
the deformation parameter $q$ is directly related to the statistics
$\nu$ of the anyon oscillators by $q=\exp({\rm i}\pi\nu)$.
\vfill
\eject
\centerline{\bf 5. Conclusions}
\vskip 1cm
We conclude this paper with a few comments.
If $\nu=1$, {\it i.e.} $q=-1$, the anyonic oscillators
have ordinary bosonic statistics and no brading phases appear
in their commutation relations. Therefore one should expect
that the Schwinger construction in this case yields a standard
Lie algebra. This is precisely what happens, because $SU(2)_{-1}$
is equivalent to the non-compact Lie algebra $SU(1,1)$ [30]. It is
worthwhile to stress that even though the oscillators have bosonic
statistics when
$\nu=1$, they are not ordinary bosons because they satisfy a
hard-core condition, or equivalently a Pauli exclusion
principle
(cf eq. (3.48a)),  like the fermions from which they originate
via the Jordan-Wigner transformation. As a matter of fact, for
any value of $\nu$, such a hard-core constraint is an essential
ingredient of our construction since it is heavily used in showing
that the generators made out of anyons actually close the standard
quantum group commutators. Indeed the proof of the equivalence between
eqs. (4.16a) and (4.17) is based on the fact that in any site ${\bf x}$
of the lattice, the operator $J^0({\bf x})$ admits only the eigenvalues
0 and $\pm 1/2$. Therefore, for our purposes  it is essential to use
fermion based anyonic oscillators.

It is well known that the Schwinger construction of $SU(2)$ can be
easily generalized to
$SU(N)$ by using $N$ sets of oscillators instead of two. Thus, we
expect that our construction of $SU(2)_q$ can be extended to $SU(N)_q$
with no difficulty [31].
It could be interesting also to extend our construction to the case
in which anyons are defined on a continuum space instead of a lattice.
In such a case, one should replace all discrete sums with suitably
defined integrals both in the disorder operators and, more generally,
in the definition of the comultiplication. Finally, it seems even
more interesting to find dynamical systems of anyons in 2+1
dimensions (either on a lattice or in the continuum) which are
endowed with this quantum group symmetry, and in particular to
study the physical consequence of this rich algebraic structure.

\vskip 3cm
\centerline{\bf Acknowledgments}
\vskip 1cm
\noindent
We would like to thank L. Castellani for many useful
discussions.

\noindent
S. Sc. would like to thank the Institute for Theoretical
Physics of S.U.N.Y. at Stony Brook for the kind hospitality
extended to him during the early stages of this work.
\vfill
\eject
\centerline{\bf References}
\vskip 1cm
\item{[1]}J.M. Leinaas and J. Myrheim, {\it Nuovo Cim.}
{\bf 37B} (1977) 1;
\medskip
\item{[2]}F. Wilczek, {\it Phys. Rev. Lett.}
{\bf 48} (1982) 114;
\medskip
\item{[3]}F. Wilczek, in {\it Fractional Statistics
and Anyon Superconductivity} edited by F. Wilczek
(World Scientific Publishing Co., Singapore 1990);
\medskip
\item{[4]}A. Lerda, {\it Anyons: Quantum Mechanics of Particles
with Fractional Statistics} (Springer-Verlag, Berlin, Germany 1992);
\medskip
\item{[5]}For a review see for example {\it The Quantum Hall
Effect} edited by R.E. Prange and S.M. Girvin (Springer-Verlag,
Berlin, Germany 1990);
\medskip
\item{[6]}Y.-S. Wu, {\it Phys. Rev. Lett.} {\bf 52} (1984)
2103;
\medskip
\item{[7]}V.G. Drinfeld, {\it Sov. Math. Dokl.} {\bf 32}
(1985) 254; {\bf 36} (1988) 212;
\medskip
\item{[8]}M. Jimbo, {\it Lett. Math. Phys.} {\bf 10}
(1985) 63; {\bf 11} (1986)247;
\medskip
\item{[9]}N. Yu. Reshetikhin, L.A. Takhtadzhyan and
L.D. Faddeev, {\it Leningrad Math. J.} {\bf 1} (1990) 193;
\medskip
\item{[10]}L. Alvarez-Gaum\'{e}, C. G\'{o}mez and
G. Sierra, {\it Nucl. Phys.} {\bf B330} (1990) 374;
\medskip
\item{[11]}V. Pasquier and H. Saleur, {\it Nucl. Phys.}
{\bf B330} (1990) 523;
\medskip
\item{[12]}D. Bernard and G. Felder, {\it Nucl. Phys.}
{\bf B365} (1991) 98;
\medskip
\item{[13]}E. Celeghini, in  {\it Symmetries in Science VI} edited by
B. Gruber (Plenum, New York, NY, USA 1992) and references therein.
\medskip
\item{[14]}J. Schwinger, in {\it Quantum Theory of
Angular Momentum} edited by L.C. Biedenharn and
H. van Dam (Academic Press, New York, NY, USA 1965);
\medskip
\item{[15]}A. Macfarlane, {\it Journ. Phys.} {\bf A22}
(1989) 4581;
\medskip
\item{[16]}L.C. Biedenharn, {\it Journ. Phys.} {\bf A22}
(1989) L873;
\medskip
\item{[17]}T. Hayashi, {\it Comm. Math. Phys.} {\bf 127}
(1990) 129;
\medskip
\item{[18]}O.W. Greenberg, {\it Phys. Rev. Lett.} {\bf 64} (1990) 705;
\medskip
\item{[19]}J. Fr\"{o}hlich and P.A. Marchetti,
{\it Comm. Math. Phys.} {\bf 112} (1987) 343;
\medskip
\item{[20]}P. Jordan and E.P. Wigner, {\it Z. Phys.}
{\bf 47} (1928) 631;
\medskip
\item{[21]} E. Fradkin, {\it Phys. Rev. Lett.}
{\bf 63} (1989) 322;
\medskip
\item{[22]}M. L\"{u}scher, {\it Nucl. Phys.}
{\bf B326} (1989) 557;
\medskip
\item{[23]}V.F. M\"{u}ller, {\it Z. Phys.}
{\bf C47} (1990) 301;
\medskip
\item{[24]}D. Eliezer and G.W. Semenoff,
{\it Phys. Lett.} {\bf 266B} (1991) 375;
\medskip
\item{[25]}D. Eliezer, G.W. Semenoff and S.S.C. Wu,
{\it Mod. Phys. Lett.} {\bf A7} (1992) 513;
\medskip
\item{[26]}D. Eliezer and G.W. Semenoff,
{\it Ann. Phys.} {\bf 217} (1992) 66;
\medskip
\item{[27]}R. Jackiw and S.-Y. Pi, {\it Phys. Rev.}
{\bf D42} (1990) 3500;
\medskip
\item{[28]}E. Fradkin, {\it Field Theories of
Condensed Matter Systems} (Addison-Wesley, Reading, MA, USA 1991);
\medskip
\item{[29]}E. Fradkin and L.P. Kadanoff,
{\it Nucl. Phys.} {\bf B170} (1981) 1;
\medskip
\item{[30]}T.L. Curtright and C.K. Zachos, {\it Phys. Lett.} {\bf 243B}
(1990) 237;
\medskip
\item{[31]}R. Caracciolo and M. R.- Monteiro,
in preparation.
\medskip

\vfill\eject
\centerline{\bf Figure Captions}
\vskip 1cm
\item{Fig. 1}{Representation of the elementary lattice
angle $\varphi({\bf x},{\bf y}^\star,{\bf x}+{\hat{\bf 2}})$
under which the link between ${\bf x}$ and ${\bf x}+{\hat{\bf 2}}$
is seen from the point ${\bf y}^\star$ of the dual lattice.}
\bigskip
\item{Fig. 2}{The elementary plaquette $A_x$ whose lower
left corner is the point ${\bf x}$.}
\bigskip
\item{Fig. 3}{Representation of the lattice angle
$\Theta_{{\cal P}_x}({\bf x},{\bf y})$
between the base point $B$ and ${\bf x}$ measured
along the curve ${\cal P}_x$ from the point
${\bf y}^\star$.}
\bigskip
\item{Fig. 4}{Examples of the curves ${\cal C}_x$
for a few points of the lattice.}
\bigskip
\item{Fig. 5}{The angle between ${\bf x}$ and
${\bf x}+\hat{\bf 1}+\hat{\bf 2}$ centered in ${\bf y}^\star$
is equal to the angle between the lines $({\bf x}\,{\bf y}^\star)$
and $({\bf x}^\star\,{\bf y})$.}
\bigskip
\item{Fig. 6}{Examples of the cuts $\gamma_x$ for a few points
of the lattice.}
\bigskip
\item{Fig. 7}{The elementary plaquette $\tilde{A}_x$ whose
upper right corner is the point ${\bf x}$.}
\bigskip
\item{Fig. 8}{Examples of the curves ${\cal D}_x$ for a
few points of the lattice.}
\bigskip
\item{Fig. 9}{Examples of the cuts $\delta_x$ for a few
points of the lattice.}
\bigskip
\item{Fig. 10}{Exchanging trajectories for type ${\cal C}$
anyons. In order not to cross the cut in ${\bf x}$, the
particle in ${\bf y}$ has to move counterclockwise if ${\bf x}>{\bf y}$
and clockwise if ${\bf x}<{\bf y}$.}
\bigskip
\item{Fig. 11}{Exchanging trajectories for type ${\cal D}$ anyons.
In order not to cross the cut in ${\bf x}$, the particle in ${\bf y}$
has to move clockwise if ${\bf x}>{\bf y}$ and counterclockwise if
${\bf x}<{\bf y}$. Notice that ${\bf x}>{\bf y}$ means
${\bf x}_{\cal C}>{\bf y}_{\cal C}
\Longleftrightarrow{\bf x}_{\cal D}<{\bf y}_{\cal D}$
(cf eq. (3.38)).}

\vfill\eject

\nopagenumbers
\hskip 9cm \vbox{\hbox{DFTT 73/92}
\hbox{ITP-SB-92-73}
\hbox{December 1992}}
\vskip 1.5cm
\centerline{{\bf ANYONS AND QUANTUM GROUPS}~
\footnote{$^*$}{
Work supported in part by Ministero dell' Universit\`a e
della Ricerca
Scientifica e Tecnologica, and by NSF grant PHY 90-08936.}}
\vskip 0.6cm
\centerline{{\bf Alberto Lerda}\footnote{$^1$}{
Also at
{\sl Institute for Theoretical Physics, S.U.N.Y.
at Stony Brook,
Stony Brook, N.Y. 11794, U.S.A.}
} ~~and~~ {\bf Stefano Sciuto}}
\vskip 0.3cm
\centerline{\sl Dipartimento di Fisica Teorica}
\centerline{\sl Universit\'a di Torino, and I.N.F.N.
Sezione di Torino}
\centerline{\sl Via P. Giuria 1, I-10125 Torino, Italy}
\vskip 2.5cm
\centerline{{\bf Abstract}}
\vskip 0.6cm
\noindent
Anyonic oscillators with fractional statistics are built
on a two-dimensional square lattice by means of a generalized
Jordan-Wigner construction, and their deformed commutation
relations are thoroughly discussed. Such anyonic oscillators,
which are non-local objects that must not be confused with
$q$-oscillators, are then combined \`a la Schwinger to
construct the generators of the quantum group $SU(2)_q$
with $q=\exp({\rm i}\pi\nu)$, where $\nu$ is the anyonic
statistical parameter.
\bye